# Glassy freezing of orbital dynamics in $FeCr_2S_4$ and $FeSc_2S_4$

R. Fichtl [a], P. Lunkenheimer [a,*], J. Hemberger [a], V. Tsurkan [a,b], A. Loidl [a]

[a] *Experimental Physics V, Center for Electronic Correlations and Magnetism, University of Augsburg, 86135 Augsburg, Germany*
[b] *Institute of Applied Physics, Academy of Sciences of Moldova, Chisinau, R. Moldova*

____________________________________________________________________________________________________

**Abstract**

We report on a thorough dielectric investigation of the glass-like freezing of the orbital reorientation-dynamics, recently found for the crystalline sulpho-spinels $FeCr_2S_4$ and $FeSc_2S_4$. As the orbital reorientations are coupled to a rearrangement of the surrounding ionic lattice via the Jahn-Teller effect, the freezing of the orbital moments is revealed by a relaxational behaviour of the complex dielectric permittivity. Additional conductivity (both dc and ac) and contact contributions showing up in the spectra are taken into account by an equivalent circuit description. The orbital relaxation dynamics continuously slows down over six decades in time, before at the lowest temperatures the glass transition becomes suppressed by quantum tunnelling.

____________________________________________________________________________________________________

## 1. Introduction

Aside of the canonical glass-forming materials, where glassy freezing suppresses long-range structural order, in recent years glassy dynamics was also investigated in crystalline systems revealing glass-like behaviour with respect to different degrees of freedom. Most prominent among those are the spin glasses where disorder-derived frustration suppresses any long-range magnetic order and the moments cooperatively freeze into a glassy low-temperature state [1]. Further examples are the so-called plastic crystals and the orientational glasses, where the centres of mass of the molecules form a regular crystalline lattice but the molecules are dynamically disordered with respect to their orientational degrees of freedom [2,3]. Very recently, based mainly on specific heat experiments, a new type of glassy freezing was reported for certain sulpho-spinels, namely the continuous slowing down of the reorientational dynamics of electronic orbitals, triggered by geometrical frustration [4,5].

Dielectric spectroscopy is a well-established technique for the investigation of the glassy dynamics in supercooled liquids, orientational glasses, and plastic crystals, giving valuable information concerning the relaxational behaviour and the many-decades change of time scales during glassy freezing (see, e.g., refs. [2,3,6,7]). Thus it is suggestive to employ dielectric spectroscopy also for the investigation of the liquid- or even glass-like orbital dynamics, suggested for $FeCr_2S_4$ and $FeSc_2S_4$ to arise from geometrical frustration effects [4,5]. Of course, a prerequisite for the applicability of this technique is the coupling of the electrical field to the orbital degrees of freedom. In Jahn-Teller active systems as those investigated here, such a coupling may arise from the fact that the orbital reorientations are coupled to the elastic response of the ionic lattice via electron-phonon interactions and thus any orbital reorientation is accompanied by a redistribution of charges. In $FeCr_2S_4$ and $FeSc_2S_4$, the lower e-doublet of the $Fe^{2+}$ ion is occupied by three electrons and hence Jahn-Teller active [4,5]. As the $Fe^{2+}$ ion is tetrahedrally coordinated by the sulphur ions, it seems reasonable that orbital reorientations are dielectrically active due to the accompanying rearrangement of the distortion of the tetrahedron. Indeed a first dielectric investigation of the orbital dynamics in $FeCr_2S_4$ at $T < 20$ K revealed a distinct relaxation-like behaviour, which can be ascribed to the freezing of orbital degrees of freedom [8]. Here we provide further results on $FeCr_2S_4$ and present relaxation-like behaviour due to orbital freezing also for $FeSc_2S_4$. For both systems, the mean relaxation time is determined, revealing glass-like slowing down, however the actual glass-transition being suppressed by quantum-mechanical fluctuations.

____________________________________________________________________________________________________

* Corresponding author. Tel. +49 821 598 3603; fax: +49 821 598 3649; e-mail: peter.lunkenheimer@physik.uni-augsburg.de

## 2. Experimental procedures

Sample preparation and subsequent heat treatments of the poly- and single-crystalline samples are described in ref. [5]. A standard four-point technique was employed to determine the dc conductivity. For the dielectric measurements, silver paint contacts were applied to the plate-like samples forming a parallel-plate capacitor. The conductivity and permittivity were measured over a frequency range of more than ten decades (0.1 Hz < $\nu$ < 2 GHz) at temperatures down to 1.4 K. A frequency-response analyzer was used for frequencies $\nu$ < 1 MHz and a reflectometric technique employing an impedance analyzer at $\nu$ > 1 MHz. For further experimental details the reader is referred to Ref. [9].

## 3. Results

Figure 1 shows the dielectric constant $\varepsilon'$ and the conductivity $\sigma'$ of single-crystalline $FeCr_2S_4$ for various frequencies at temperatures below 80 K. $\varepsilon'(T)$ shows a strong step-like decrease with decreasing temperature, exhibiting characteristic relaxational behaviour [7]. These relaxation steps shift towards low temperatures with decreasing frequency. For high temperatures and/or low frequencies, $\varepsilon'$ reaches a colossal plateau value of about 8000. In addition, a further increase of $\varepsilon'$ is observed at the lowest frequencies investigated (only partly shown), approaching values up to $10^7$. The relaxation steps of $\varepsilon'$ are accompanied by peaks or shoulders in $\sigma'(T)$. Having in mind that $\sigma' \sim \varepsilon'' \times \nu$, corresponding features will show up in the dielectric loss, too. Dependent on the measuring frequency, above about 25 K the measured $\sigma'$ is smaller than the four-point dc conductivity $\sigma_{dc}$, given by the line in Fig. 1(b). As indicated by the arrows, at low temperatures another shoulder shows up in $\sigma'(T)$, shifting towards higher temperatures with increasing frequency. In $\varepsilon'(T)$ [Fig. 1(a)] a corresponding small relaxation step can be suspected to show up at values of $\varepsilon' < 10^2$, being superimposed to the strong steps reaching colossal values, mentioned above. The inset of Fig. 1, comparing the dielectric loss $\varepsilon''(T)$ of a single- and a poly-crystalline sample, demonstrates that the shoulder is absent in the polycrystalline sample, $\varepsilon''$ being more than one decade smaller at the peak temperature of the corresponding single-crystal curve.

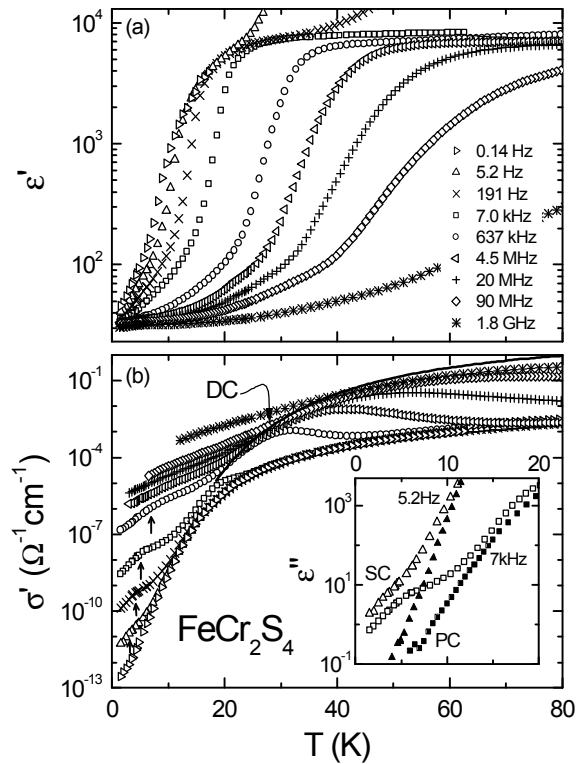

Fig. 1. Temperature dependence of the dielectric constant (a) and conductivity (b) in single-crystalline $FeCr_2S_4$ for selected frequencies. The line in (b) is the result of a four-point dc measurement; the arrows indicate the position of the intrinsic relaxation loss peaks connected to orbital freezing. Inset: Comparison of the dielectric loss in single- and polycrystalline $FeCr_2S_4$ at two frequencies.

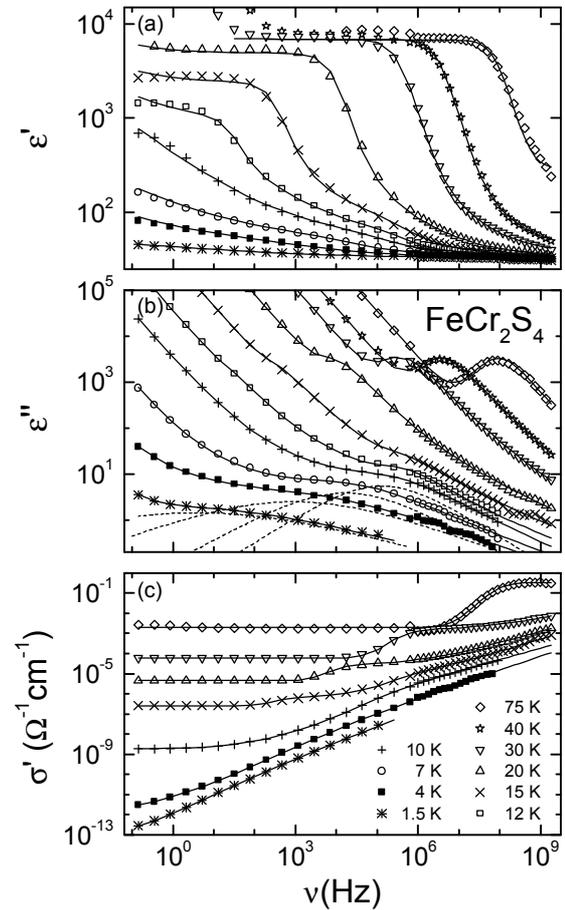

Fig. 2. Dielectric constant (a), loss (b) and conductivity (c) of single-crystalline $FeCr_2S_4$ vs. frequency at various temperatures (to keep the figure readable, in (c) not all temperatures are shown). The solid lines represent the results of least-square fits with an equivalent circuit as described in the text. For the lowest temperatures, the dashed lines in (b) characterise the contributions due to the orbital relaxation.

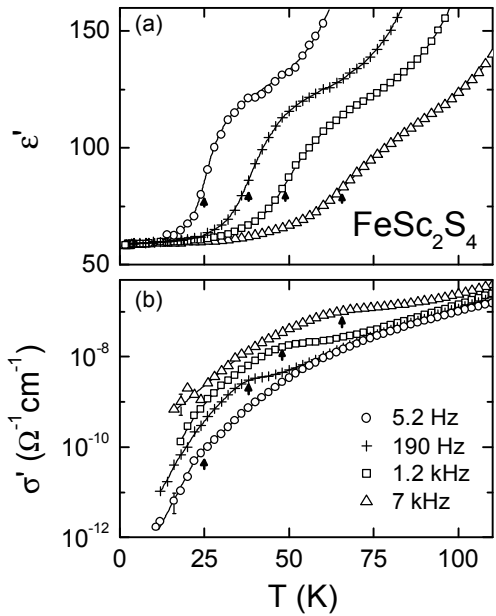

Fig. 3. Temperature dependence of the dielectric constant (a) and conductivity (b) in single-crystalline FeSc$_2$S$_4$ for selected frequencies. The arrows indicate the position of the relaxational feature, connected to orbital freezing. The lines are drawn to guide the eyes.

Figure 2 shows the frequency-dependent dielectric constant, loss and conductivity of single-crystalline FeCr$_2$S$_4$ for selected temperatures. $\varepsilon'$ is dominated by sigmoidally shaped steps, typical for relaxation processes [7]. In accordance with Fig. 1(a) it reaches colossal values at low frequencies. $\varepsilon''(\nu)$ exhibits a divergence towards low frequencies, indicating strong conductivity contributions, which is in accord with the constant $\sigma'(\nu)$ observed at low-frequencies in Fig. 2(c). In addition, two relaxation features show up as loss peaks or shoulders in Fig. 2(b), with an amplitude of several 1000 or of about 10, respectively. As shown in Fig. 2(c), the high-amplitude loss peaks correspond to well-pronounced upward steps of $\sigma'(\nu)$.

First results on single-crystalline FeSc$_2$S$_4$, being isostructural to FeCr$_2$S$_4$, are given in Fig. 3. It shows the temperature dependence of the dielectric constant and conductivity for selected frequencies. As indicated by the arrows, in this material, being isostructural to FeCr$_2$S$_4$, a clear signature of relaxational behaviour shows up, too. Well-developed shoulders in $\sigma'(T)$ are accompanied by a step-like decrease towards low frequencies in $\varepsilon'(T)$, both features simultaneously shifting towards higher temperatures with increasing frequency. The high-temperature plateau of the relaxation steps in Fig. 3(a) is not fully developed but superimposed by a further increase of $\varepsilon'(T)$.

## 4. Discussion

The strong relaxation feature observed in FeCr$_2$S$_4$ (Figs. 1 and 2), accompanied by colossal values of $\varepsilon'$, can be completely ascribed to non-intrinsic contact effects [10,11]. Metal-semiconductor contacts often exhibit a high ohmic resistance, accompanied by a high capacitance due to the formation of a depletion layer at the metal-to-semiconductor interface. They usually can be modelled by a parallel RC-circuit, connected in series to the bulk sample [10,11,13]. This leads to a so-called Maxwell-Wagner relaxation, with a frequency and temperature dependence that can easily be mistaken for an intrinsic relaxational behaviour [11]. At high frequencies, the contact resistor becomes shortened by the contact capacitance and the intrinsic bulk response is detected. In FeCr$_2$S$_4$, this causes the smooth step-like increase of $\sigma'(\nu)$, observed at the higher temperatures in Fig. 2(c). Indeed, the values reached by $\sigma'$ beyond the step agree well with the results from the four-point dc measurement, which are not affected by contact contributions [compare Fig. 1(b)]. The corresponding features in the complex permittivity are the high-amplitude peak in $\varepsilon''$ [Fig. 2(b)] and the strong relaxation step in $\varepsilon'$ [Fig. 2(a)]. The observed shift of these features with temperature (Fig. 2) respectively frequency (Fig. 1), can be understood taking into account the temperature-dependence of the circuit's time constant, which is mainly determined by the semiconducting characteristics of the sample resistance. The apparently colossal values of the dielectric constant in FeCr$_2$S$_4$ are caused by the high capacitance of the thin depletion layer. For a more detailed discussion of contact contributions in dielectric measurements and their analysis, see e.g. [11].

As already mentioned, at sufficiently high frequencies (or low temperatures) the contact contributions no longer play a role and the intrinsic bulk response is detected. For $FeCr_2S_4$ the four-point dc conductivity, included in Fig. 1(b) enables a simple determination of the intrinsic region: For all temperatures and frequencies where $\sigma'$ is smaller than $\sigma_{dc}$, contact contributions cannot be neglected. In contrast, whenever $\sigma'$ is larger than $\sigma_{dc}$ the intrinsic response of the sample is detected. Thus the relaxational features indicated by the arrows in Fig. 1(b) clearly are of intrinsic character. The observed shoulders can be ascribed to loss peaks, superimposed to a contribution due to charge transport. The latter is composed of the dc conductivity (below 15 K approximately represented by the 0.14 Hz curve) and an ac conductivity increasing with frequency, which has a weaker temperature dependence than $\sigma_{dc}$ and can be ascribed to hopping transport of localized charge carriers [12]. The relaxational behaviour occurs in just the same temperature region, where evidence for glassy freezing of the orbital dynamics was deduced from specific heat measurements, revealing a smeared-out peak in $C_p/T$ vs. T at about 5 K [5]. Thus the detected dielectric relaxation process can be assumed to mirror the glass-like slowing down of orbital dynamics. Further evidence for this notion arises from a comparison to the results on a polycrystalline sample shown in the inset of Fig. 1(b). In polycrystalline samples the orbital degrees of freedom were found to be ordered at low temperatures, most likely due to impurity effects [5]. While for both samples the conductivity background is nearly identical, the relaxation feature is completely missing in the polycrystal. This finding strongly corroborates the view of orbital reorientations leading to the observed relaxation feature.

The frequency-dependent response shown in Fig. 2 was fitted employing an equivalent circuit assuming a parallel RC circuit for the contacts and an additive combination of dc- and ac conductivity with a Cole-Cole function for the bulk response. For the ac conductivity a power law, $\sigma' \sim \nu^s$ with s < 1 was used. It corresponds to Jonscher´s "Universal Dielectric Response" [13], which usually is ascribed to hopping conduction of localized charge carriers [12]. The phenomenological Cole-Cole function [7,14] is often employed to parameterize loss peaks in orientational glasses [2]. Good agreement of fits (solid lines in Fig. 2) and experimental spectra could be achieved in this way. Only the strong additional increase of $\varepsilon'$ towards values up to $10^7$, evident at the two highest temperatures in Fig. 2(a) (only partly shown), is not taken into account by this circuit. The origin of this behaviour is unclear up to now; possibly it is due to an additional ac conductivity of the contact resistor. The dashed lines in Fig. 2(b) show the intrinsic relaxational parts of the fits. Very broad loss peaks are revealed. For the renowned Debye case, where an identical exponential time-dependence for all relaxing entities is assumed, loss peaks with a half-width of 1.14 decades are expected, which is clearly exceeded in the present case. Thus the orbital relaxation in $FeCr_2S_4$ shows the typical broadening, characteristic for glassy systems, which is commonly ascribed to a disorder-induced heterogeneous distribution of relaxation times [7,15]. With decreasing temperature the peak width increases significantly, while its amplitude decreases. Such behaviour is commonly observed in the so-called orientational glasses, crystalline materials where the molecules exhibit frustration-induced disorder with respect to the orientational degrees of freedom [2]. It can be explained assuming a temperature-independent Gaussian distribution of energy barriers hindering the reorientational motion [16].

In $FeSc_2S_4$ (Fig. 3) no apparently colossal values of $\varepsilon'$ show up, indicating that the complete response is of intrinsic nature. Obviously, contact contributions can be neglected in this material due to the much lower conductivity compared to $FeCr_2S_4$ as revealed by comparing Figs. 1(b) and 3(b). Thus in Fig. 2, the clear signature of intrinsic relaxational behaviour shows up. Well-developed shoulders in $\varepsilon''$, indicating an underlying loss peak superimposed to charge transport contributions as in $FeCr_2S_4$, are accompanied by a step-like increase towards high temperatures in $\varepsilon'$. These steps do not fully saturate, but exhibit a further increase at high temperatures. This can be ascribed to ac-conductivity, leading to a contribution also in $\varepsilon'$ [12,13]. We conclude that also in $FeSc_2S_4$ the orbital dynamics shows a glass-like slowing down towards low temperatures, which leads to the observed relaxation features. This again is in good accord with the finding of a broad peak in the low-temperature specific heat of this material [4], characteristic for glassy freezing. In addition, in Ref. [17], just in the temperature region where we observe glassy dynamics in our experiments, a broadening of Mössbauer lines was observed for $FeSc_2S_4$, which can be taken as indication of a slowed-down orbital motion.

The most significant parameter characterizing glassy freezing is the temperature development of the relaxation time $\tau$. As revealed in Fig. 4, for both materials investigated the relaxation time characterizing the reorientational dynamics of the orbitals shows a continuous slowing down over a broad range (up to six decades) with decreasing temperature, which is typical for glassy freezing. In the Arrhenius representation of Fig. 4, a purely thermally activated process corresponds to a straight line. At the higher temperatures the curves in Fig. 4 indeed may be approximated by Arrhenius behaviour as is indicated for $FeSc_2S_4$ by the solid line corresponding to an energy barrier of 27 meV (inset). However, in contrast to most other glassy systems, the temperature dependence of $\tau$ becomes weaker for low temperatures, for $FeCr_2S_4$ even seeming to level off at a constant value of about 0.1 s. A constant $\tau(T)$ can be explained by tunnelling processes driving the orbital reorientations and hence we interpret the overall behaviour as a smooth transition from a thermally activated to tunnelling-dominated dynamics towards low temperatures. As already evident from a comparison of Figs. 1 and 3, the temperature range of the glassy slowing down of orbital motion in $FeSc_2S_4$ is higher than in $FeCr_2S_4$. Thus, while the origin of this difference is unresolved up to now, it seems reasonable that thermally activated behaviour dominates in this material, tunnelling processes leading to small deviations below about 30 K only (inset of Fig. 4). It should be noted that, especially for $FeCr_2S_4$, the onset of tunnelling may prevent the relaxation time to become larger than 100 s, a limit where for canonical glass formers often the glass transition is defined. Therefore, strictly spoken, within this conventional definition of the glass transition the actual orbital glass state may not be realized in these materials. However, of course relaxation times in the 0.1 – 1 s region as observed here at low temperatures, certainly are extremely slow for electronic degrees of freedom.

and relaxation steps in the plots of $\sigma'(T)$ and $\varepsilon'(T)$, respectively (cf. arrows in Fig. 3). The inset gives a magnified view of $\tau(1/T)$ of $FeSc_2S_4$. The dashed lines are guides to the eyes. The solid line in the inset indicates thermally activated behaviour with an energy barrier of 27 meV.

## 5. Conclusions

In conclusion, our results show that dielectric spectroscopy is also suited to investigate the recently discovered glassy freezing of the reorientational motion of electronic orbitals. Via the Jahn-Teller effect, the orbital reorientations are dielectrically active and lead to characteristic relaxational features in the dielectric response of $FeCr_2S_4$ and $FeSc_2S_4$ at low temperatures. We find typical glassy behaviour, in particular a continuous slowing down of the orbital dynamics and a broad distribution of relaxation times. A complete freezing-in is suppressed by quantum-mechanical tunnelling, limiting the low-temperature relaxation time in $FeCr_2S_4$ to about 0.1 s.

## Acknowledgements

This work was supported by the Deutsche Forschungsgemeinschaft via the Sonderforschungsbereich 484 and partly by the BMBF via VDI/EKM, FKZ 13N6917.

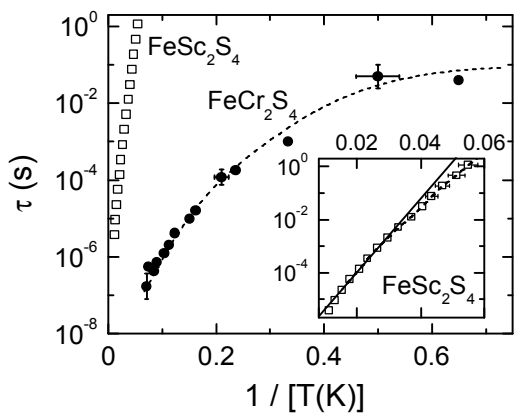

Fig. 4. Temperature dependence of the relaxation time of $FeCr_2S_4$ and $FeSc_2S_4$ in Arrhenius representation. For $FeCr_2S_4$, $\tau$ was obtained from the fits as shown in Fig. 2. For $FeSc_2S_4$ it was estimated from the positions of the shoulders